%%%%THIS NEEDS PHYZZX MACROS%%%%
\input phyzzx
\def\Bbb #1{\hbox{\bf #1}}
%%%% If you do not have msam and msbm fonts, comment out the
%%%% following  line.
\font\Bb=msbm10 \def\Bbb #1{\hbox{\Bb #1}}
\hsize=15.8cm
\vsize=23cm

\newwrite\ffile\global\newcount\figno \global\figno=1
\def\fig{fig.~\the\figno\nfig}
\def\nfig#1{\xdef#1{fig.~\the\figno}%
\writedef{#1\leftbracket fig.\noexpand~\the\figno}%
\ifnum\figno=1\immediate\openout\ffile=figs.tmp\fi\chardef\wfile=\ffil
e%
\immediate\write\ffile{\noexpand\medskip\noexpand\item{Fig.\
\the\figno. }
\reflabeL{#1\hskip.55in}\pctsign}\global\advance\figno by1\findarg}

\parindent 25pt
\overfullrule=0pt
\tolerance=10000
\def\RR{${\rm R}\otimes{\rm R}~$}
\def\NSNS{${\rm NS}\otimes{\rm NS}~$}

\def\Tr{{\rm Tr}}
\def\half{{\textstyle {1 \over 2}}}

\nopagenumbers
\baselineskip=14pt
\REF\duffa{M. Duff and R.  Khuri, {\it String Solitons},
hep-th/9412184,
  Phys. Rep.  {\bf 259}  (1994) 213.}
\REF\schwarzx{J.H.  Schwarz, {\it   an $SL(2,Z)$ multiplet of type
IIB
 superstrings},  hep-th/9508143,  Phys.  Lett. {\bf B360} (1995) 13.}
\REF\polchina{J. Polchinski, {\it Dirichlet-branes and Ramond--Ramond
charges},
hep-th/9510017, Phys. Rev. Lett. {\bf 75} (1995) 4724.}
\REF\wittenv{J. Polchinski  and E.  Witten, {\it Evidence for
heterotic -- type
I duality}, hep-th/9510169.}
\REF\bergshoeffa{E.  Bergshoeff, M.B.  Green, G.  Papadopoulos and
P.K.
Townsend, {\it  The IIA super eightbrane},
hep-th/9511079.}
\REF\polchinb{J.  Polchinski and A.  Strominger, {\it New vacua for
type II
string theory}, hep-th/9510227.}
\REF\romansa{L.  Romans, {\it Massive N=2a supergravity in ten
dimensions},
Phys.  Lett. {\bf 169B} (1986) 374.}
\REF\polchinf{J.  Polchinski, {\it  Combinatorics of boundaries in
string
theory},  hep-th/940731, Phys.  Rev. {\bf D50} (1994) 6041.}
\REF\greena{M.B.  Green, {\it A gas of $D$-instantons},
hep-th/9504108,
Phys.Lett. {\bf B354} (1995) 271.}
\REF\greenh{M.B. Green, {\it Point-like states for type IIB
superstrings},
Phys.  Lett. {\bf B329} (1994) 435.}
\REF\Einstein{ A. Einstein and N. Rosen,
 {\it The particle problem in the general theory of relativity}
Phys. Rev. {\bf 48} (1935)  73.}
\REF\greenea{B.R.  Greene, A.  Shapere, C.  Vafa and  S-T.  Yau, {\it
Stringy
Cosmic Strings}, Nucl.  Phys.  {\bf B337}  (1990)  1.}
\REF\Schwarz{J H Schwarz, {\it Covariant Field Equations of Chiral
 N=2 D=10 Supergravity} Nucl Phys {\bf B226} (1983) 269.}
\REF\hulla{C.M.  Hull and P.K.  Townsend,
 {\it Unity of superstring dualities},
 hep-th/9410167, Nucl. Phys. {\bf 438} (1995) 109.}
\REF\wittena{ E.  Witten, {\it String theory dynamics
 in various dimensions},    hep-th/9503124,
 Nucl.  Phys. {\bf B443} (1995) 85.}
\REF\shenkera{S.  Shenker,
 {\it The strength of non-perturbative effects in string theory},
 Proceedings of the  Cargese Workshop on Random Surfaces,
 Quantum Gravity and Strings, Cargese, France, May 28 - Jun 1, 1990.}
\REF\Comtet{ A. Comtet and G W. Gibbons,
 {\it Bogomol'nyi Bounds for Cosmic Strings},
 Nucl. Phys {\bf B299} (1988) 719.}
\REF\nepomechie {R I Nepomechie, {\it Magnetic monopoles
 from antisymmetric tensor gauge fields}
Phys. Rev. {\bf D31} (1985) 1921.}
\REF\teitelboima {C. Teitelboim,
{\it Gauge invariance for extended objects} Phys. Lett. {\bf B167}
(1986) 63.}
\REF\teitelboimb{C. Teitelboim,
 {\it Monopoles of higher rank} Phys. Lett. {\bf B167} (1986) 69.}
\REF\Rey{ S-J Rey, {\it The confining phase of superstrings
 and axionic strings}
 Phys Rev {\bf D43} (1989) 526.}
\REF\giddingsa{S B Giddings and A Strominger,
 {\it String wormholes} Phys Lett {\bf B230}
(1989) 46.}
\REF\dufflu{ M.J. Duff and X.Lu, {\it Remarks on String/Five-brane
duality}
 Nucl. Phys. {\bf B354} (1991) 129.}
\REF\Hawkinga{ S W Hawking, {\it Breakdown of predictability
in gravitational collapse} { Phys Rev} {\bf D14} (1976) 2460.}
\REF\Hawkingb{ S W Hawking, {\it Wormholes in Spacetime}
Phys Rev {\bf D37} (1986) 904}
\REF\Gross {D Gross, {\it Is quantum gravity unpredictable?}
 Nucl Phys {\bf B236} (1984) 349.}
\REF\Coleman{ S. Coleman, {\it Black Holes as Red herrings}
Nucl Phys {\bf B207} (1988) 867-887.}
\REF\Colemanb{ S Coleman and S Hughes,
 {\it Black Holes, Wormholes and the Disappearance of Global Charge}
 Phys Lett {\bf B309}  (1993) 246.}
\REF\hawkingc{S.W.  Hawking, {\it Virtual Black Holes},
hep-th/9510029.}

\hfill {DAMTP/R95-56}\break
 \null
\vskip 3cm
\centerline{\bf INSTANTONS AND SEVEN-BRANES}
\centerline{\bf IN TYPE IIB SUPERSTRING THEORY.}
\vskip 1cm
 \centerline{ Gary W. Gibbons, Michael B.  Green and Malcolm J.
Perry}
\vskip 0.2cm
\centerline{DAMTP, Silver Street, Cambridge CB3 9EW, UK.
\foot{G.W.Gibbons@damtp.cam.ac.uk; M.B.Green@damtp.cam.ac.uk;
malcolm@damtp.cam.ac.uk}}
\nopagenumbers
\vskip 2cm
Instanton and seven-brane solutions of type IIB supergravity
 carrying charges in
the Ramond-Ramond sector are constructed.  The singular seven-brane
has a
quantized \RR \ \lq magnetic'  charge whereas its  dual is
the  instanton, which is non-singular in the string frame and
 has an associated
{\it global}
\lq electric' charge.  The product of these charges is constrained by
a Dirac
quantization condition.  The instanton  has the form of a space-time
wormhole in the string frame, and  is responsible for the
non-conservation of the Noether current.

\vfill\eject
\pagenumbers
\sequentialequations

\chapter{\bf   Introduction}
Recent developments in our understanding of superstring theory
suggest that all
known theories may be viewed as different perturbative approximations
of a single
 underlying theory.  Viewed from the standpoint of any given string
theory
the fundamental strings of other theories are solitonic states that
are not
apparent in string perturbation theory.  Other $p$-branes also arise
as
solitons, suggesting an important r\^ole for objects with all
possible values of $p$ in the theory.   In type II theories the
fundamental
strings carry the charges of the Neveu--Schwarz-Neveu--Schwarz
(\NSNS)
sector but
do not carry Ramond--Ramond (\RR) charges.  These latter charges
 are associated with
$(p+1)$-form potentials or $(p+2)$-form field strengths, $F_{p+2}$,
that are
carried by some of the known $p$-branes of the these theories.  The
solitons
that carry the \RR charge that have been constructed so far are the
 zero-brane (black hole),
two-brane, four-brane and six-brane solutions of the type IIA theory
and the
one-brane
(string), self-dual three-brane and five-brane solitons of type IIB
string
theory (for a review see [\duffa]).  In fact, there is an
infinite-dimensional $SL(2,{\Bbb Z})$ multiplet of both
\lq dyonic'  one-branes [\schwarzx] and five-branes in the type IIB
theory.

Recently,  some $p$-brane solitons  have been associated with
superstring
configurations known generically as $D$-branes  (or $D$-instantons in
the
 $p=-1$
case) [\polchina] which are sources of the \RR charge.  This
association
suggests that there should be solitons for {\it all} values of $p$
from $p=-1$
to $p=9$.  The case $p=9$ is very special since the accompanying
field strength
vanishes identically, and is connected with the presence of chiral
anomalies in
type I theories with any gauge group other than $SO(32)$. The $p=8$
soliton
constructed in [\wittenv\bergshoeffa] couples to a cosmological
constant in the
 type IIA
theory [\polchina, \polchinb] and is a solution of \lq massive' type
IIA
supergravity [\romansa] .  This paper considers the   $p=-1$
(instanton)
and $p=7$ (seven-brane) solutions of the type IIB theory which have
\RR
charges that are related by a Dirac-like quantization condition.
$D$-instantons were previously considered in the bosonic theory in
[\polchinf,\greena] and the BPS boundary condition for the type IIB
theory
$D$-instanton was obtained in [\greenh].

Although the construction to be described bears a resemblance to the
construction of the previously discovered $p$-branes, there are
fascinating new
features.  For example, the
instanton solution is non-singular in the string frame -- in fact
it is a wormhole.
Whereas other $p$-brane solutions can be thought of as wormholes with
infinitely
long
throats the instanton is genuinely an Einstein-Rosen wormhole
[\Einstein]
which connects
two asymptotically euclidean regions of space-time.
The \lq electric' charge carried by the instanton is an \RR charge
that flows through the wormhole throat, and is interpreted as
 a violation of the conservation of a
global charge in physical processes.
 The seven-brane solution,
carrying the dual \lq magnetic' charge, is related to the
stringy cosmic string
solution of [\greenea].

The solutions of   type IIB supergravity  we will consider are ones
in which
 the
two scalar fields (the dilaton, $\phi$, and the \RR scalar, $a$)  and
the
metric have non-trivial behaviour while the other bosonic fields (the
two
third-rank field strengths and the self-dual fifth-rank field
strength) vanish.
 The ten-dimensional  lagrangian for the non-vanishing fields is
$${\cal L} = R - \half  (\partial\phi)^2- \half e^{2\phi}   (\partial
a)^2
\eqn\einstein$$
in the Einstein frame (where the signature is $(- + + + + + + + +\
+)$).
Defining a nine-form field strength, $F_{(9)} = e^{2\phi} *da$, i.e.,
$$F_{\mu_1 \dots \mu_9} = e^{2\phi} \epsilon^\mu_{\ \mu_1 \dots
\mu_9}\partial_\mu a  ,\eqn\nineform$$
the lagrangian can be written in the equivalent form,
$${\hat {\cal L}} = R  - \half   (\partial\phi)^2 - {1\over 2 (9!)}
e^{-2\phi}
F_{\mu_1 \dots \mu_9}F^{\mu_1 \dots \mu_9}.\eqn\dualact$$
The passage from \einstein\ to \dualact\ is a standard duality
transformation.
The field equation for $F_{(9)}$ coming from \dualact~\  is
$$\nabla_\mu \left(e^{-2\phi} F^\mu_{\ \mu_1\dots
\mu_8}\right)=0,\eqn\noeth$$
which is equivalent to the Bianchi identity for $a$.

\chapter{\bf  The instanton solution}
The  form of $\hat{\cal L}$  remains unchanged after Wick rotation to
euclidean
signature.
The field equations that arise from the euclidean version of
\dualact\ have a
form that could have been obtained from
$$\tilde{\cal L} = R - \half (\partial \phi)^2 + \half e^{2\phi}
(\partial
\alpha)^2,
\eqn\eucagain$$
which is  \einstein\  with the substitution $a \to \alpha =  i a$.
The
integrals of the Lagrangians
 \dualact\ and \eucagain\ give actions that differ purely
 by surface terms that can arise in
replacing $da$ by $F_{(9)}$.  The  euclidean equations of motion
are invariant under a euclidean
version of  $N=2$ supersymmetry.   The equations of motion that
follow from
\eucagain\ are
$$\eqalign{R_{\mu\nu} - \half \left(\partial_\mu \phi \partial_\nu
\phi -
e^{2\phi} \partial_\mu \alpha  \partial_\nu \alpha\right) &= 0 , \cr
\nabla_\mu \left(e^{2\phi} \partial ^\mu \alpha\right) & =0,\cr
\nabla^2\phi - e^{2\phi} (\partial \alpha)^2 & =0.}\eqn\equations$$

We now turn to the conditions that need to be satisfied for a
solution of the
euclidean theory to  preserve half the total supersymmetries which
requires a
brief discussion of the euclidean version of the $N=2$ supersymmetry
of the
type IIB theory.  This could be expressed in terms of symplectic
spinors  but
it is more succinct to express the euclidean theory in a manner that
parallels
the usual discussion with lorentzian signature.   Using the
conventions in
[\Schwarz] the lorentzian signature supersymmetry transformations of
the
fields can be written in a complex notation.  They can then be
systematically
adapted to the case of euclidean signature by replacing the usual
algebra
${\Bbb C}$ of complex numbers, generated over the reals ${\Bbb R}$ by
$1$ and
$i$ (where  $i^2=-1$) by the so-called \lq double' or hyperbolic
complex
numbers ${\Bbb E}$, which are generated over the reals by $1$ and
$e$,  where
$e^2= +1$.  Conjugation of complex numbers is an automorphism taking
$1$ to $1$
and $i$ to $-i$.  Similarly conjugation of double numbers takes $1$
to $1$ and
$e$ to $-e$.  Thus a general double number $u=a + eb$ ($a,b \in {\Bbb
R}$) is
conjugate to $u^* = a-eb$ and $uu^*=a^2-b^2$.  As an algebra over
${\Bbb R}$
the double numbers are reducible ${\Bbb E} = {\Bbb R}\oplus {\Bbb
R}$. The
associated projectors $P_{\pm} = \half (1\pm e)$ are lightlike,
$P_{\pm}
P^*_{\pm} =0$, with respect to the indefinite inner product $uu^*$.

The supersymmetry transformations of the ${\Bbb C}$-valued spin-$1/2$
and
spin-$3/2$ fields, $\lambda$ and $\psi_\mu$  are given in  [\Schwarz]
(in the
Einstein frame).   In a bosonic background in which only
 $\tau=a+ie^{-\phi}=\tau_1+i\tau_2$ and
$g_{\mu\nu}$ are non-zero they are,
$$\delta \lambda = -{1\over\tau_2}\left({\tau^*-i\over\tau+i}\right)
\gamma^\mu
 (\partial_\mu\tau_1 + i \partial_\mu \tau_2)
(\epsilon_1-i\epsilon_2),\eqn\lamtrans$$
$$\delta \psi_\mu = \left(\partial_\mu + {1\over 4}
\omega_\mu^{ab}\gamma_a\gamma_b - i{1\over2} Q_\mu\right)
(\epsilon_1+ i\epsilon_2),
\eqn\psitrans$$
where the composite $U(1)$ gauge potential is defined by
$$Q_\mu = {-1\over 4\tau_2}\left\{ \left({\tau-i \over
\tau^*-i}\right)
(\partial_\mu\tau_1 - i\partial_\mu\tau_2) + \left({\tau^* + i \over
\tau+i}\right) (\partial_\mu \tau_1 + i\partial_\mu \tau_2) \right\}
\eqn\qdef$$
These expressions  have been transformed from the $SU(1,1)$-invariant
form in
[\Schwarz]  to the parameterization with manifest $SL(2,\Bbb
R)$-invariance.

For the application to the  euclidean instanton solution, we change
$i$ to $e$
everywhere in these transformations and  make the identifications
$\tau_1 =
\alpha$ and $\tau_2 =  e^{-\phi}$.  The ansatz that leads to the
preservation
of half the supersymmetry is $d\tau_1 = \pm d\tau_2$ and the metric
is flat, i.e.
$$d\alpha = \pm e^{-\phi}d\phi,\qquad g_{\mu\nu} =
\delta_{\mu\nu}.\eqn\supans$$
It follows from \lamtrans\ that   $\delta \lambda =0$ if $\epsilon_1
=\pm
\epsilon_2$.   From hereon we will arbitrarily choose the plus sign.
Furthermore, to obtain $\delta\psi_\mu=0$ we note that the spin
connection
vanishes ($\omega_\mu^{ab}=0$) since the metric is flat in the
Einstein frame.
Thus, after using the ansatz  $\tau_1 = \tau_2 + k$ (with constant
$k$) the
transformation of the gravitino becomes,
$$\delta\psi_\mu = (1+e)\left (\partial_\mu
+{k^2 -1\over4\tau_2((\tau_2+k)^2 - (1+ \tau_2)^2)}
\right)\epsilon_1.\eqn\grvitrans$$
This can be made to vanish by setting $\epsilon_1 =
f(\tau_2)\epsilon_1^0$,
where $\epsilon_1^0$ is an arbitrary constant real spinor,  and
choosing
$f(\tau_2)$ appropriately.

The conditions \supans\  combined with the equations of motion
\equations\ lead
to
$\partial^2\phi = -  (\partial \phi)^2$,
so that
$$\partial^2 (e^{\phi}) =0.\eqn\harmon$$
This  has a   spherically symmetric solution describing a single
instanton,
$$e^{\phi}  = \left(e^{\phi_\infty}  + {c\over
r^8}\right),\eqn\phisol$$
where $\phi_\infty$ is the value of the dilaton field  at $r=\infty$
and $c$ is
a constant that is arbitrary at this stage, and will be shown below
to be
proportional to the instanton charge. From \supans~ $\alpha$ is given
by
$$ \alpha - \alpha_\infty =- e^{-\phi} + e^{-\phi_\infty}
\eqn\defalpha $$
where $\alpha_\infty$ is the constant value of $\alpha$ at
$r=\infty$.
The solution is specified by the value of the two constants
$\phi_\infty$ and $\alpha_\infty$.

This single instanton solution is evidently singular at $r=0$ in the
Einstein
frame.  However, it is natural to transform from the Einstein frame
(in which the
metric is simply $ds_E^2 =  dx^2$)  to the string frame in which the
metric is
given by,
$$ds^2 = e^{\phi/2} ds_E^2 =   \left(e^{\phi_\infty} + {c\over
r^8}\right)^{1/2} \left(dr^2 + r^2
d\Omega_9^2\right),\eqn\stringmet$$
where $d\Omega_n^2$ is the $SO(n)$-invariant line element on $S^n$.
This metric is
manifestly invariant under the inversion transformation,
$$r \to (ce^{-\phi_\infty} )^{1/4}{1\over r},\eqn\invert$$
which shows that the region $r\rightarrow 0$ is another
asymptotically
Euclidean region identical to that near $r=\infty$.
In fact, the solution in this
frame is a wormhole in which there are two asymptotically euclidean
regions
connected by a neck.  The space-time is geodesically complete so in
this sense
it is non-singular.

The euclidean action of the instanton is given (in the Einstein
frame), by
$$ I_{inst} =   \int_M  \left(-R + \half(\partial \phi)^2   + {1\over
2
(9!)} e^{2\phi} F_{\mu_1\dots \mu_9} F^{\mu_1\dots \mu_9}\right)  -
2\int_{\partial M} [\Tr  K],\eqn\eucact$$
where $[\Tr  K]$ is the difference between the trace of the extrinsic
curvature
on the boundary and the value it would have if the boundary were
filled in with
flat space.  For our instanton solution   $R=0$ and  the boundary
contributions
also vanish. The contribution from infinity clearly vanishes. That
from
$r=0$ also vanishes since $\Tr K={9\over r}$, but when integrated
over
a small nine-sphere of radius $r$, it will vanish.
     Using the fact that
$(\partial\phi)^2= e^{-2\phi}F_{\mu_1\dots \mu_9} F^{\mu_1\dots
\mu_9} /9!$,
one obtains
$$I_{inst}= \int_{R^{10}} (\partial\phi)^2 = - \int_{R^{10}}
\partial^2\phi \ = -\int_{r=\infty} \partial_\mu \phi \ d\Sigma^\mu +
\int_{r=0} \partial_\mu \phi\ d\Sigma^\mu.\eqn\actaga$$
Using the explicit form of $\phi$ in \phisol\ the contribution from
$r=0$ is
seen to vanish whereas the contribution from $r=\infty$ gives the
total
action,
$$ I_{inst} = {8c \over g}~ {\rm Vol}~(S^9) = {2\pi^{5/2}\over 3}
c,\eqn\totalactwo$$
where
${\rm Vol}(S^9)=2\pi^{5/2} /\Gamma(5)$ is the volume of the unit
nine-sphere
and  where $g=e^{\phi_\infty} $ is the string coupling constant.

An electric charge $Q^{(-1)}$ is defined by the Noether charge
for the translation symmetry $\alpha \rightarrow \alpha + {\rm
constant}$.
 Thus
$$ Q^{(-1)}= \int_{S^9}e^{2\phi}F^\mu d\Sigma_\mu =
\int_{S^9} J_\mu^{Noether} d\Sigma^\mu = \int_{S^9}
F_{(9)}\eqn\eleccharge $$
where $F_\mu=\partial_\mu\alpha$,
$J_\mu^{Noether}=e^{2\phi}\partial_\mu\alpha$, and the integration is
over the nine sphere at $r=\infty$.
Conservation of the Noether current is equivalent to the field
equation
for $F_{(1)}$,
$$ \nabla^\mu J_\mu^{Noether} = \nabla_\mu(e^{2\phi}F^\mu)=0
\eqn\morelectcharge $$
For our solution
$$ Q^{(-1)} = \int e^{2\phi}\partial_\mu\alpha d\Sigma^\mu =
\int e^\phi \partial_\mu\phi d\Sigma^\mu = -8 c g~{\rm Vol}(S^9)
\eqn\evenmorecharge $$
so that
$$ I_{inst}= {\vert Q^{(-1)}\vert\over g}.\eqn\ratio$$
This shows the
expected dependence of the instanton action on the coupling constant.
The
energy of the $p$-brane solitons of the \RR sector have a similar
$1/g$
dependence [\hulla,\wittena].
 Such effects are expected from general considerations
in string theory [\shenkera]  based on estimates of the divergence of
closed-string perturbation theory and matrix model calculations.
Furthermore,
we will show later that  the constant $c$ is determined by a
quantization
condition that fixes $Q^{(-1)}$ to be $2\pi n$, where $n$ is an
integer and the
action \ratio\  is  that of a D-instanton  [\polchina]. The charge
$Q^{(-1)}$ has a clear interpretation in the string frame as the
charge
that flows down the throat of the wormhole.
It thus represents the violation of the charge in a physical
 spacetime process.

Multi-instanton solutions, with $k+1$ asymptotically Euclidean
regions
located at $x_i$  are obtained by taking the solution
of \harmon,
$$e^{\phi} = e^{\phi_\infty} +\sum_{i=1}^{i=k}
 {c_i\over (x - x_i)^8},\eqn\phisoltwo$$
where $c_i$ are positive constants that define the
 charges,  $\vert Q^{(-1)}_i\vert
 =  8c_i g~ {\rm Vol}(S^9)$. All of these instantons are simply
connected.

\chapter{\bf The seven-brane solution}
The seven-brane is a soliton with an eight-dimensional world-volume
characterized by non-trivial behaviour of the scalar fields and the
metric in
the two transverse euclidean dimensions.
Defining $\tau_1 = a$ and $\tau_2 = e^{-\phi}$ (where $\tau = \tau_1
+ i\tau_2$
spans the upper-half plane) the lagrangian \einstein\ is identified
with  the
$SL(2,\Bbb R)$-invariant lagrangian considered in [\greenea],
$$I = \int_M  \left(-R +  {(\partial\tau_1)^2+ (\partial\tau_2)^2
\over2 \tau_2^2} \right)- \int_{\partial M} 2 [\Tr  K],
\eqn\sevenact$$
The global  $SL(2,\Bbb R)$ symmetry of this action is expected to be
broken in
string theory to a residual $SL(2,\Bbb Z)$ that
 is interpreted as a local symmetry,
so that  $\tau$  should be  restricted to the fundamental domain of
the modular
group, $|\tau|\ge 1$, $-1/2 \le \tau_1\le 1/2$. Thus, the target
space of the
sigma model is a non-compact orbifold of finite volume with two
orbifold
points at $\tau=i$ and $\tau=e^{i\pi\over3}$, and a cusp at
$\tau_2=\infty$.
Many arguments have been
advanced  that point to this breaking of  the continuous global
symmetry.
This restriction of the domain of the scalar fields is an essential
ingredient
in the following. Moreover it appears that unless one restricts the
domain
in this way, there are probably no finite energy solutions.

The ansatz for the seven-brane solution is one in which the fields
are trivial
in the eight dimensions of the world-volume of the brane and
non-trivial in the
transverse space ($r$, $\theta$) so that the metric in the Einstein
frame takes
the form,
$$ds_E^2 = -dt^2 + (dx^1)^2 + \dots (dx^7)^2 + \Omega^2 (dr^2 +
r^2d\theta^2),
\eqn\metseven$$
where $\Omega=\Omega(r,\theta)$. Introducing the complex co-ordinate
$z=re^{i\theta}$,
the complex scalar field $\tau$ will be taken to satisfy the
holomorphic (or
antiholomorphic) ansatz, $\bar \partial  \tau =0$ (or $\partial \tau
=0$).
 The Einstein
 equation for
$\Omega$ is
$$ \partial\bar\partial\ln\Omega =
{2\partial\tau\bar\partial\bar\tau\over
(\tau-\bar\tau)^2} = 2\partial\bar\partial\ln\tau_2. \eqn\einseven$$
Just as in the case of the  stringy cosmic string of [\greenea]~
the   single seven-brane  solution  is obtained by choosing
$\tau(z)$ so that the pull-back of the elliptic modular
function has a single pole,
for example a pole at infinity,
$$ j(\tau(z)) = b z,\eqn\modeqn$$
The constant $b$ determines the value of the dilaton as $r\to
\infty$.
In that case  $\Omega\sim r^{-{1\over12}}$
as $r\rightarrow\infty$ and the space transverse to the seven-brane
is
 asymptotically conical with deficit angle $\delta={\pi\over6}$.
We can easily see that this is consistent with supersymmetry by
considering
\lamtrans\ and \psitrans.
We will now use with complex numbers to describe supersymmetry,
rather
than double numbers, because
the signature is Lorentzian.  Demanding that
$\gamma^1\gamma^2(\epsilon_1 + i \epsilon_2) =\pm i(\epsilon_1+
i\epsilon_2)$
enforces the condition $\delta \psi_\mu=0$ since
 $\omega_\mu^{12} = Q_\mu$
for this ansatz.   In other words  the spin connection and the
composite gauge
connection cancel.  Asymptotically conical spacetimes usually do not
admit
covariantly constant spinors, but in this case, the non-trivial
gravitational
holonomy  is cancelled by that of the $U(1)$ gauge field.
The fact that $\delta \lambda =0$ follows from \lamtrans\ making
use of the holomorphicity of the field $\tau$.
%$\tau_1$ and $\tau_2$ which implies,
%$$\gamma^1\partial_1\tau_1 + i\gamma^2\partial_2 \tau_2 =
%\gamma^1\partial_1\tau_1 - i\gamma^2\partial_1 \tau_1 ,\eqn\holone$$
%$$\gamma^2\partial_2\tau_1 + i\gamma^1\partial_1 \tau_2 =
%\gamma^2\partial_1\tau_2 + i\gamma^1\partial_1 \tau_2 .\eqn\holone$$

A solution of the equations of motion that follows from \sevenact\
has $R =
\half((\partial  \tau_1)^2 + (\partial\tau_2)^2)/\tau_2^2$ so that
the energy
of the solution comes entirely from the boundary contribution. The
energy per
unit seven-volume of the seven-brane with these boundary conditions
satisfies
a Bogomol'nyi bound
$$ E=2\delta \geq {\pi\over3}, \eqn\bogbound$$
with equality if and only if $\tau$ is either holomorphic or
antiholomorphic
corresponding to the supersymmetric case [\Comtet].

The magnetic charge $P^{(7)}$ of the seven-brane is
$$ P^{(7)} = \oint F_\mu d\Sigma^\mu =
-{1\over2\pi}\int_0^{2\pi}d\theta = -1 \eqn\magcharge $$
where the line integral is taken around a closed loop at infinity,
and we have used the fact that as $r\rightarrow\infty,
 a \sim -{\theta\over 2\pi}$.
This charge is localized on the inverse images of $i$ and
$e^{i\pi\over3}$,
the two orbifold points of the target space. Thus this seven-brane
behaves as
one would expect of a singular  source for the magnetic field.

One can straightforwardly extend
the analysis of [\greenea]~ for multi-strings
 to find multi-seven-brane solutions
by replacing \modeqn \ by
$$ j(\tau(z)) = {P(z)\over Q(z)} \eqn\ratmodeqn $$
where $P(z)$ and $Q(z)$ are polynomials in $z$ of order $m$ and $n$
respectively with no common factors. If $m>n,$ we obtain
a solution with $k=m$ seven-branes. If $m\leq n,$ the solution
represents $k=n$ seven-branes.
 Provided that $k<12$, the transverse space is asymptotically
conical. If $k=12$, the transverse space is asymptotically
cylindrical,
and if $k>12$ the transverse space has finite volume; in  general it
is
singular apart from the exceptional case of $k=24$. There appears to
be an
interesting difference between multi-seven-branes and
multi-instantons.
 Two seven-branes may continuously approach one another until they
coincide,
the resultant seven-brane having a charge equal to the sum of the
individual
 charges. By contrast, for our instantons the analogous process does
not
 seem to
be possible.

 \chapter{\bf Quantization condition}
The \RR charges carried  by the seven-brane and the instanton are
related by
the same quantization condition that applies to magnetic and electric
charges
of the other pairs of branes with $p$ and $\tilde  p$
 [\nepomechie,\teitelboima,\teitelboimb] where in
the case of superstrings $p+\tilde p =6$,
$$P^{(\tilde p)} Q^{(p)} = 2\pi n, \qquad n\in \Bbb Z.\eqn\quantcon$$
In our case, $p =-1$ and $\tilde p=7$,  so that  substituting the
fact that the
electric charge of the elementary seven-brane solution  has a
quantized charge,
$P^{(7)} = m$ (where $\vert m\vert =0,1,\dots, 11,12,24)$, gives a
quantization
condition on the charge carried by the instanton,
$$Q^{(-1)} =2\pi n.\eqn\instval$$
Using \evenmorecharge\ this determines $c$
 in the single  instanton solution  to
be
$$c= {3\vert n\vert \over \pi^{3/2}}.\eqn\cdeter$$
With this value the action for the instanton in \ratio\  agrees  with
that of  the $D$-instanton [\polchina], where it is determined in an
altogether
different  procedure as a functional integral over a string
world-sheet with
the topology of a disk.

\chapter{\bf Discussion}

In this paper we have argued that the D-Instanton of type IIB string
theory
may be identified with a  BPS spacetime wormhole  solution
of the  euclidean ten-dimensional type IIB supergravity theory
carrying  \RR electric charge.
  We have also argued that the dual magnetically charged seven-brane
may be identified with a solution constructed from the stringy
cosmic string of [\greenea]. Consistent with our interpretation,
the instanton solution has finite euclidean action
$I_{inst}= {2 \pi \over g} |Q^{(-1)}|$, where $|Q^{(-1)}|$
 is the electric charge violated by the instanton, (the charge
flowing
through the wormhole neck).
 The instanton is geodesically
complete in the string frame and the dilaton field $\phi$ is
everywhere finite
though it diverges at infinity in one  asymptotically flat region.
It is also consistent that the seven-brane  has finite energy per
unit
 seven-volume, saturates a Bogomol'nyi bound and is geodesically
complete
 in the Einstein frame,
but not in string frame. The magnetic charge resides entirely
at two point sources.
 The total magnetic charge $P^{(7)}$
is quantized and satisfies a Dirac-Teitelboim-Nepomechie quantization
condition.

Our \RR electrically charged  D-instanton or $-1$-brane
 has a number of features which distinguish it from other
 \RR  $p-$branes.
 Firstly it has a finite throat and two asymptotically
flat regions, while in  other cases any throat is necessarily
infinitely long.
This may be partly understood using supersymmetry.
The Lorentzian  solutions have Killing spinors
$\epsilon$ and the associated Killing vector
 ${\overline \epsilon} \gamma ^\mu \epsilon$ can never become
spacelike.
Thus if these solutions have regular horizons they must be of extreme
type which means that the surfaces of constant time resemble those in
the extreme Reissner-Nordstr\"om solution and have the form of
infinitely
long throats with an internal infinity rather than the finite
Einstein-Rosen throats  encountered on the surfaces of constant time
of
 non-extreme black holes. For our euclidean instanton one cannot
 construct a timelike Killing vector from  the Killing spinor.
 In fact in the string frame the metric
of our instanton coincides exactly with the constant time surfaces of
the
$11$-dimensional vacuum black hole solution of the  vacuum Einstein
 equations but we suspect that  this a coincidence.

It is also interesting to contrast our ten-dimensional \RR
 instanton with the
closely related four dimensional axionic instanton of
 the \NSNS sector [\Rey,\giddingsa].
This may also be thought of as the four dimensional
transverse space of
the neutral five-brane [\dufflu]. In string frame, the solution is
the
product of flat 6-dimensional Minkowski  spacetime
$(t, y^1,...y^5)$ with a curved four
dimensional transverse space of the form
$ds^2=  (e^{2 \phi_\infty} + { c \over r^2} )
(dr^2+r^2d\Omega_3^2)$.  In the Einstein frame
 the transverse space is flat.
If  $t$ is imaginary, the  solution may be interpreted as an
instanton
with an infinite throat.
 In the
flat four-dimensional Einstein frame the dilaton
$\phi$ and \NSNS 3-form field
strength $F_{(3)}$ satisfy the self-duality condition which
guarantees supersymmetry:
$d\phi = \pm e^{-2\phi} \ast F_{(3)}$ which is similar to the
condition
for our ten-dimensional instanton. The main difference is that due to
the \NSNS three-form $F_{(3)}$ coupling  with a different power
of $e^\phi$ compared to the \RR fields, it is the square of
$e^{\phi}$
which is  harmonic. This is why the throat is infinite rather than
finite.
Moreover, for the same reason, the Euclidean action of this  \NSNS
instanton is
proportional to $ 1 \over g^2$ not $ 1 \over g$ as it is for
 our ten-dimensional  R$\otimes$R instanton.

One of the most intriguing aspects of our work is the relation to the
breaking of continuous global symmetries. On the one hand superstring
theory is
believed to have no continuous symmetries. On the other hand, in
 low energy quantum gravity the thesis that  black holes and possibly
wormholes should lead to a violation of the conservation
 of charges associated with continuous and possibly discrete
symmetries
has been
strenuously argued [\Hawkinga,\Hawkingb], and equally strenuously
rebutted [\Gross,\Coleman].
The connection between real black holes, virtual black holes
and instantons in this context remains,
 however, obscure [\Colemanb,\hawkingc].

The
relevant continuous symmetries are contained in
$SL(2,\Bbb R)$.   From the point of view of string theory it is
believed on quite
general grounds that this  must break down to the modular subgroup
$SL(2,\Bbb Z)$.
Now $SL(2,\Bbb R) $ is certainly a  symmetry of the
classical equations of motion of
type IIB supergravity theory written in terms of the dilaton and
 pseudoscalar field $a$. In particular the equation of motion for $a$
may be
thought of as the conservation of a Noether current
$J^{\rm Noether} _\mu = e ^{ 2\phi} \partial _\mu a$ arising
 from the translation subgroup:
$ a \rightarrow a + {\rm constant} $. The associated
 charge  is $Q^{(-1)}$.
It may be argued that quantum mechanically spacetime
 wormholes will lead to the violation of the conservation
of any Noether charge
 because some of the current
 $J^{\rm Noether} _\mu = e ^{ 2\phi} \partial _\mu a$
 may flow down the throat.
It may also be argued that black holes or black branes
should lead to a violation of the conservation of Noether charges.
We shall comment on this later.

The  discussion so far has assumed that the global translational
symmetry
 $a \rightarrow a + {\rm constant}$
 is well-defined. This
would be true of our sigma model if the target space were the entire
upper half plane $SL(2,\Bbb R)/SO(2)$. But it is not.
Rather it is the fundamental domain of the modular group
$SL(2,\Bbb Z)\\SL(2,\Bbb R)/SO(2)$. Thus the translations do
not act globally on the target space of our sigma model.
In other words the continuous global symmetry is not well defined.
There are two apparently unrelated reasons for quotienting out
by the modular group. One is string inspired. The other is that
unless
we do so, we cannot obtain a seven-brane with finite total energy.

Thus both for stringy reasons and by virtue of wormhole
effects we do not expect the electric charge to be conserved. What
about the magnetic charge $P^{(7)}$ of the seven-brane and
what about black holes? The usual physical arguments for the
violation
of global charges by black holes rely on two main planks. Firstly,
there should be a no-hair theorem for the charge, and secondly
there should be a lower bound to the mass of any state carrying
the relevant charge. In our case, it is straightforward to show
that if the target space is the entire upper half plane
then there are no black hole solutions with regular event horizons
and non-constant scalar fields. If the target space is the
fundamental
domain of the modular group, the  argument is not quite so
straightforward but as far as we can tell the result seems to hold.
Moreover, the same no-hair results seem to apply to other p-branes.
 Secondly, the only states carrying \RR charge
are p-branes, in particular the only states we know of that carry
$P^{(7)}$
are the seven-branes of this paper, which satisfy a Bogomol'nyi bound
on the energy per unit seven-volume. Therefore there are no
light states in the theory that carry \RR charges.
 Thus it seems plausible that the
conservation of $P^{(7)}$ could be violated
by dropping seven-branes into a black hole or a black p-brane and
letting it evaporate.

Finally we conclude with the observation that the construction of the
D-instanton wormhole solutions in this paper opens up the prospect of
studying
a variety of non-perturbative effects of great interest in quantum
gravity
within a  {\sl controlled} computational scheme. These include
the breaking of supersymmetry, the nature of a
possible  non-perturbative dilaton potential and the problem of the
cosmological constant.

\refout
\bye